\documentclass[11pt]{article}
\usepackage{moriond,epsfig}

\def\Journal#1#2#3#4{{#1} {\bf #2}, #3 (#4)}

\def\PRL{\em Phys. Rev. Lett.}

\def\ZPA{{\em Z. Phys.} A}
\def\EJP{{\em Eur. Phys. J.} C}

\def\be{\begin{equation}}
\def\ee{\end{equation}}
\def\bea{\begin{eqnarray}}
\def\eea{\end{eqnarray}}

\begin{document}
\vspace*{4cm}
\title{FRAGMENTATION AND SPECTROSCOPY OF HADRONS IN EP COLLISIONS}

\author{N. COPPOLA, ON BEHALF OF THE H1 AND ZEUS COLLABORATIONS}

\address{Deutsches Elektronen-Synchrotron 
DESY, Notkestra\ss e 85, 22607 Hamburg, Germany}

\maketitle\abstracts{Recent results on charm fragmentation and
  hadron spectroscopy in $e^+/e^- p$ collisions at HERA with the
  ZEUS and H1 detectors are presented. The measured fragmentation
  ratios and fragmentation fractions are in agreement with those
  measured in $e^+e^-$, thus supporting the assumption of
  universality.  Measurements of the inclusive photoproduction of the
  neutral mesons $\eta, \rho^0, f_0(980)$ and $f_2(1270)$ are also
  presented. At the same time results on production of pentaquarks are
  shown. Cross sections of observed states and upper limits on the
  production cross section of unobserved states are extracted in order
  to enable comparison between experiments.}

\section{Introduction}

Understanding the process whereby quarks and gluons convert to
colour-less hadrons is one of the outstanding problems in particle
physics. 

In perturbative QCD (pQCD), the cross section for inclusive production
of a heavy hadron H can be expressed as a convolution of two terms:
\begin{equation}
\sigma(p_H)=\sum_{\mbox{\scriptsize part}} \int dz dp_{\mbox{\scriptsize part}}
\sigma(p_{\mbox{\scriptsize part}})D^{\mbox{\scriptsize part}}_H(z)\delta(p_H-z p_{\mbox{\scriptsize part}})
\end{equation}
where $\sigma(p_{\mbox{\scriptsize part}})$ is the perturbative part
of the cross section for the production of the parton and
$D^{\mbox{\scriptsize part}}_H(z)$ is the corresponding fragmentation
function. The latter contains a non-perturbative, incalculable part.
The factorisation theorem, if applicable, predicts that
$D^{\mbox{\scriptsize part}}_H(z)$ is universal, i.e. both its shape
and normalisation are independent of the hard subprocess and the scale
at which the parton was produced. This assumption needs to be verified
experimentally.

The measurements of the inclusive photoproduction of charmed mesons,
baryons and of the neutral mesons $\eta, \rho^0, f_0(980), f_2(1270)$
could also contribute to resolving the problem of the hadronisation.

Recently, some experiments have reported narrow signals in the
vicinity of 1530 MeV in the $nK^+$ and $pK_S^0$ invariant mass spectra
which are consistent with the exotic pentaquark baryon state
$\Theta^+$ with quark content $uudd\bar{s}$,~\cite{penta1} while other
experiments have searched for this state with negative results. The
possible existence of a charm pentaquark has also been discussed, with
renewed theoretical interest in calculating their expected
properties~\cite{penta2,penta3}
following the observation of strange pentaquarks.

\begin{figure}
\hspace{1.cm}
\psfig{figure=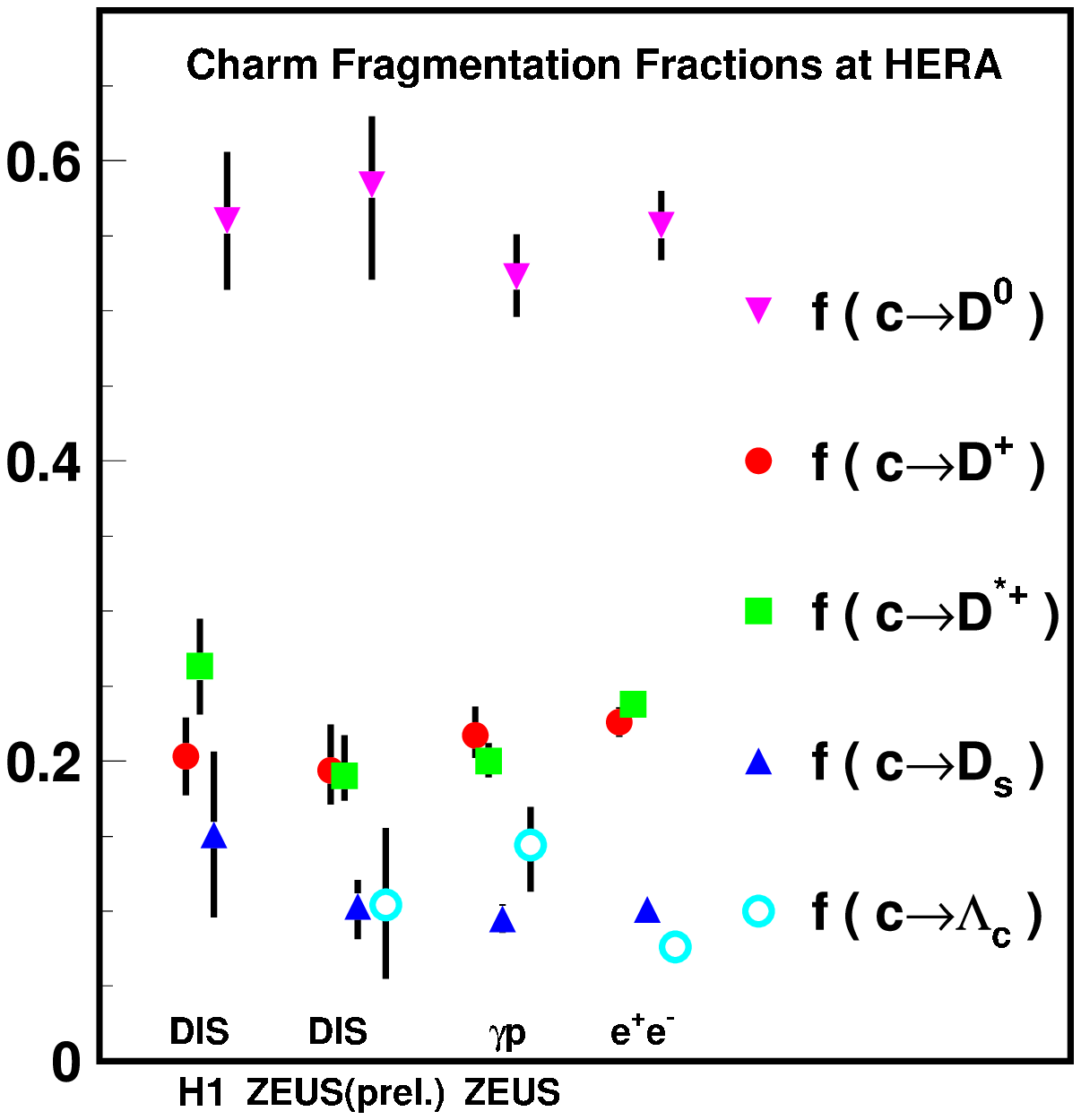,height=6.5cm}
\vspace{0.5cm}
\hspace{0.5cm}
\psfig{figure=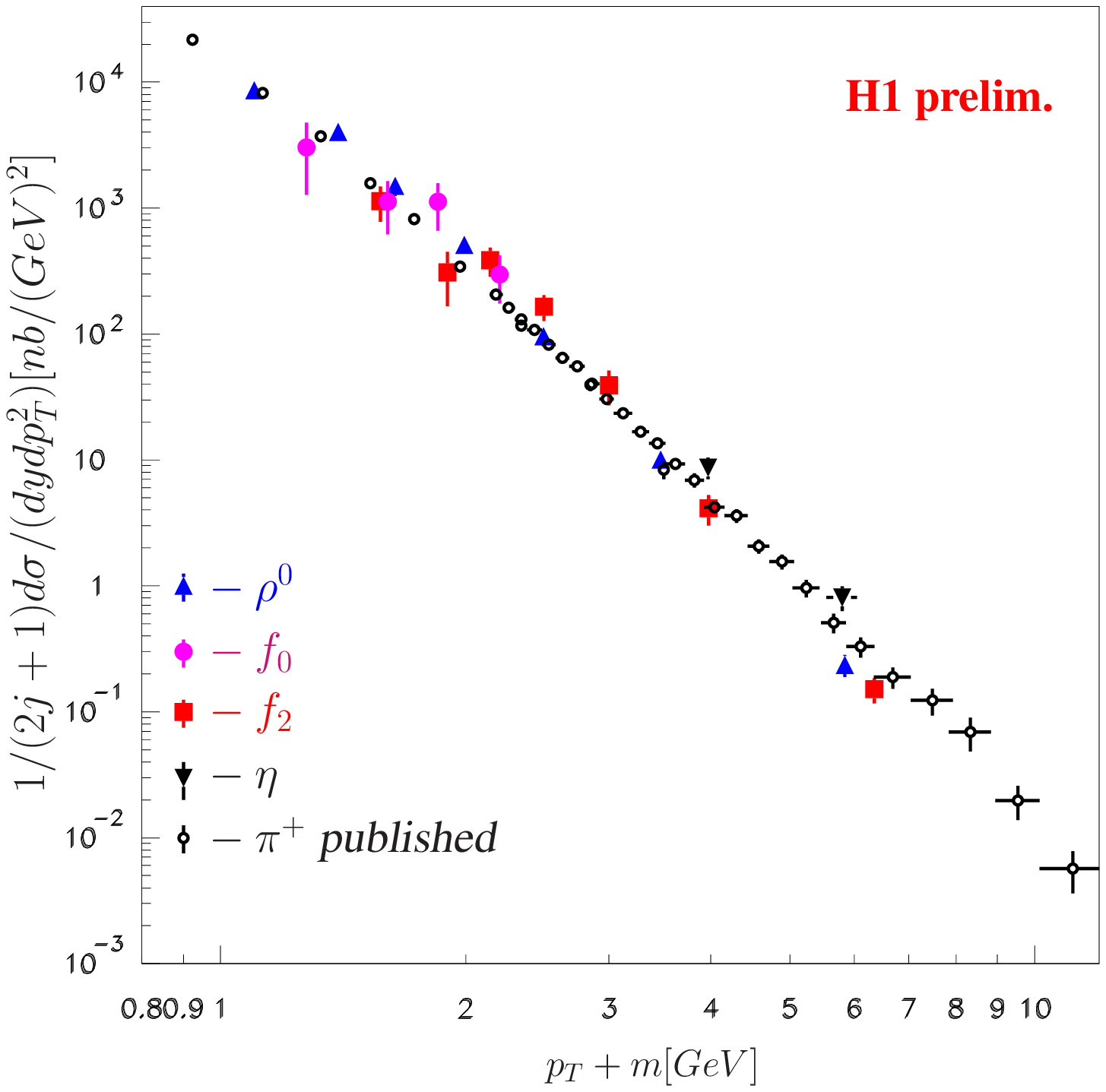,height=6.5cm}
\caption{The fragmentation fractions for charmed hadrons (see
  Sec~\ref{sec:charm}) (left). The differential photoproduction cross
  section of $\eta, \rho^0, f_0(980), f_2(1270)$) (see
  Sec~\ref{sec:light}) (right).
\label{fig:charfrag}}
\end{figure}

\section{Charm fragmentation}\label{sec:charm}

In order to study the probabilities of a heavy quark to hadronise into
various heavy hadrons, two types of observables are used. The
fragmentation fraction for a given charmed hadron is defined as the
ratio of the total production cross section for that given hadron to
that for the charm quark. Fragmentation ratios are used to highlight
certain aspects of the hadronisation process. Their exact definitions
can be found in the references~\cite{charm1,charm2,charm3}.

The ZEUS collaboration has measured the fragmentation fractions and
ratios of $D^+, D^0, D^+_s$, $D^{*+}$ and $\Lambda_c$
states\footnote{Together with their charge conjugate states} both in
deep-inelastic scattering (DIS) and
photoproduction~\cite{charm2,charm3}. These states were measured by
reconstructing the invariant mass and the number of events was determined,
after subtraction of reflections, in a fit to signal and background.
The measured cross sections are given for the visible phase space,
defined for DIS as $1.5<Q^2<1000~\mbox{GeV}^2$, $0.02<y<0.7$,
$p_t(D,\Lambda)>3~\mbox{GeV}$ and $\|\eta(D,\Lambda)\|<1.6$ and for
photoproduction as $Q^2<1~\mbox{GeV}^2$,
$130<W<300~\mbox{GeV}$,$p_t(D,\Lambda)>3.8~\mbox{GeV}$ and
$|\eta(D,\Lambda)|<1.6$. The charm quark production cross section in
the visible range, needed to calculate the fragmentation fractions,
was calculated from the measured cross sections of $D$'s and
$\Lambda_c$.

The H1 collaboration has used a different experimental procedure to
measure fragmentation ratios and fractions of  $D^+, D^0,
D^+_s,D^{*+}$ in DIS, profiting from its central silicon
tracker~\cite{charm1}. In order to improve the signal/background
ratio, cuts on the secondary vertex parameters were used. The number
of visible charmed meson states was then determined from a fit to the
invariant mass distribution. The measurement was done in the
kinematic region  $2<Q^2<100~\mbox{GeV}^2$,
$0.05<y<0.7$, $p_t(D)>2.5~\mbox{GeV}$ and $|\eta(D)|<1.5$. A
QCD-based model was used to extrapolate the measured cross sections to
the full phase space and to predict the total charm quark cross
section. The fragmentation ratios were then calculated from the
measured fragmentation fractions.

The results of H1 and ZEUS (see Fig.~\ref{fig:charfrag}(left)),
although obtained with different experimental procedures, are
compatible with each other and with the results form $e^+e^-$
experiments (with comparative errors) and thus support the assumption
of universality.

\section{Light Mesons ($\eta, \rho^0, f_0(980), f_2(1270)$)
  Photoproduction}\label{sec:light}

Besides charmed states and production of well-known hadrons such as
$\pi$, $K_S^0$, $\Lambda$, protons, and charmed mesons, $J/\psi$, {\em
  etc.} that have been measured by ZEUS and H1; a recent result is the
cross section measurement of inclusive photoproduction of $\eta,
\rho^0, f_0(980)$ and $f_2(1270)$ mesons at H1 in the central rapidity
region. In this analysis a photoproduction data sample taken in the
year 2000, corresponding to an average energy of the photon proton
centre of mass $\sqrt{s_{\gamma p}}=210~\mbox{GeV}$, was used.
The $\rho^0, f_0(980)$ and $f_2(1270)$mesons were reconstructed
through $\pi^+\pi^-$ decay , the $\eta$ meson through $\gamma\gamma$
decay. 

In Fig.~\ref{fig:charfrag}(right), the measured differential
cross section $1/(2j+1)d^2\sigma/dydp^2_T$ of the resonances as
function of $m+p_T$, where $j$ is the spin and $m$ is the mass of the
measured particle, was compared with the cross section of charged
pions. The resonances have a similar behaviour as observed for
long-lived hadrons~\cite{light1}.

\begin{figure}
\hspace{2.cm}
\psfig{figure=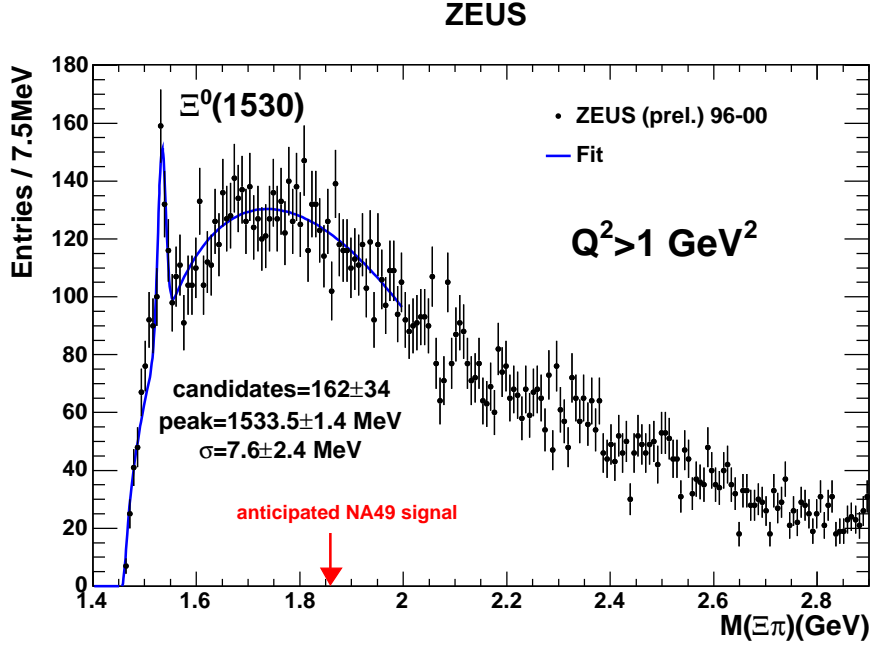,height=8.5cm}
\caption{The $\Xi\pi$ invariant mass spectrum for $Q^2>1 \mbox{GeV}^2$
  (all four charge  combinations summed).
\label{fig:xi}}
\end{figure}

\section{Pentaquarks}

\begin{figure}
\hspace{1.cm}
\psfig{figure=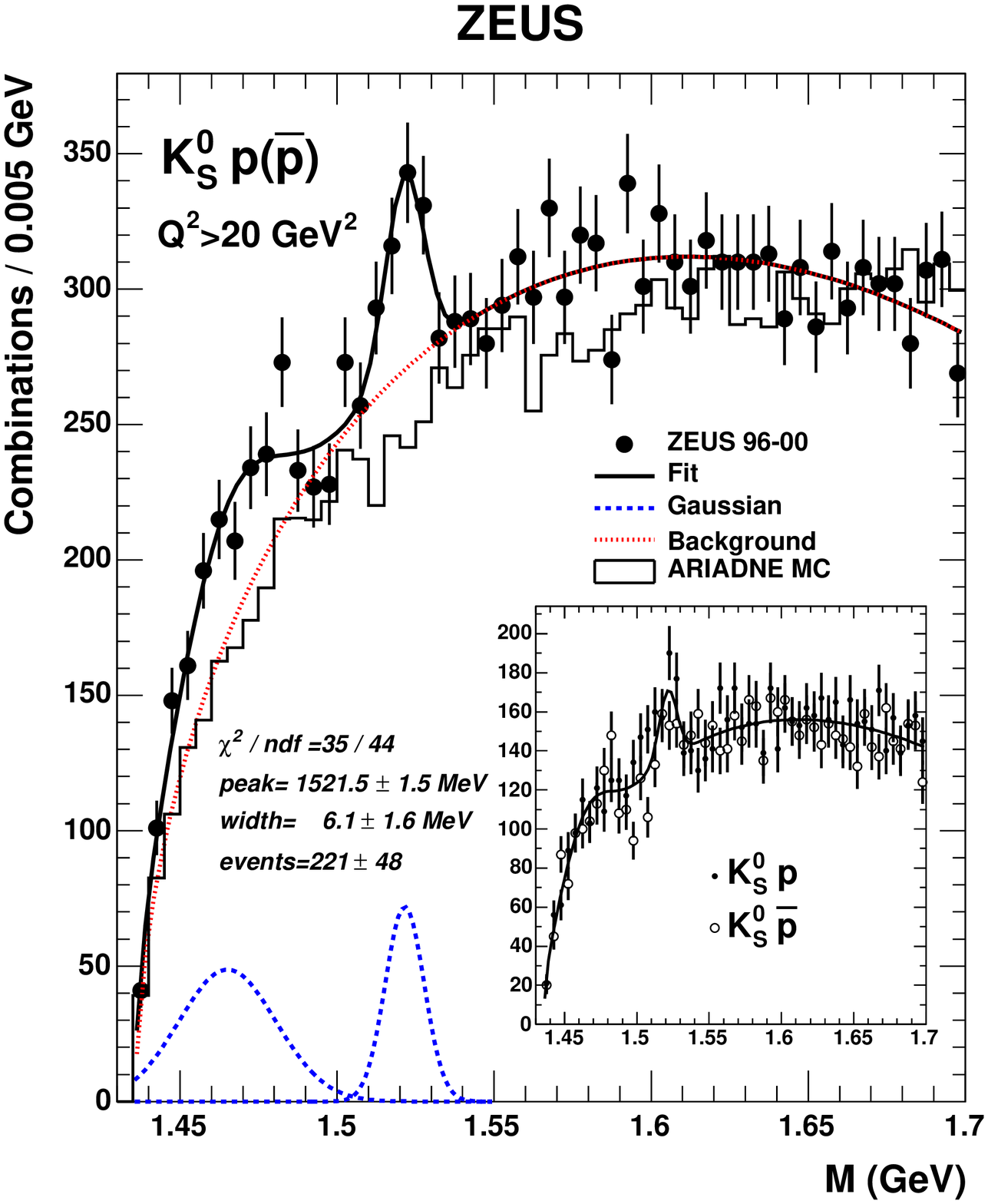,height=6.5cm}
\vspace{0.5cm}
\hspace{0.5cm}
\psfig{figure=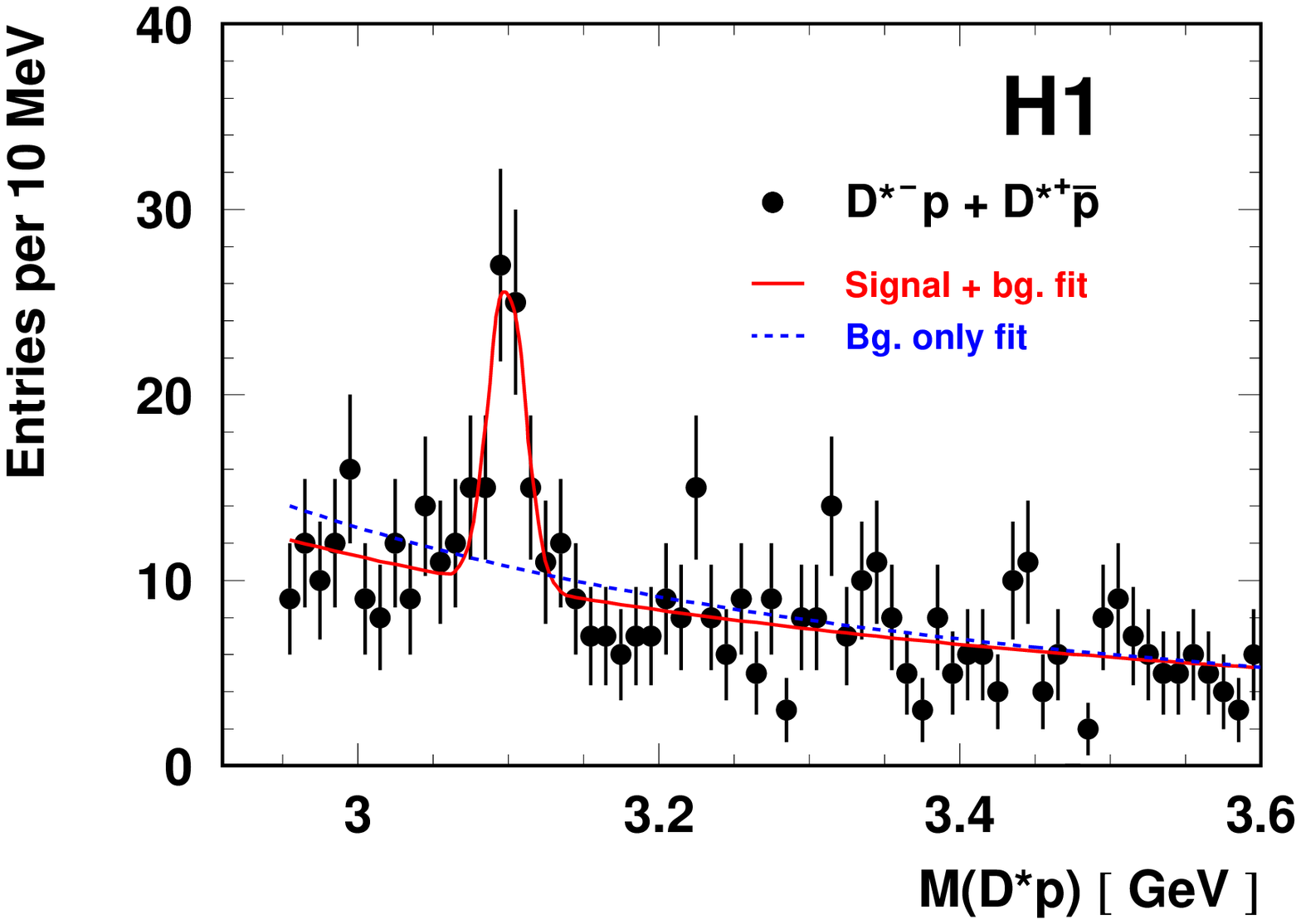,height=6.5cm}
\caption{Invariant mass spectrum for the $K_S^0p(\bar{p})$ channel for $Q^2>20~\mbox{GeV}^2$
  (left). $M(D^*p)$ distribution from opposite charge $D^*p$
  combinations in DIS. (right).
\label{fig:penta}}
\end{figure}

A strange pentaquark $\Theta^+$ candidate was seen by the ZEUS
collaboration. In this analysis deep inelastic scattering (DIS) events
were selected by requiring an exchanged photon virtuality
$Q^2>1~\mbox{GeV}^2$. The $\Theta^+$ was reconstructed in the decay
channel $pK_S^0$. The invariant mass spectrum of proton and $K^0$ for
events where $Q^2>20~\mbox{GeV}^2$ was fitted with a polynomial
background and two Gaussians. The signal peak (see
Fig~\ref{fig:penta}(left)) is observed with a mass of $1521.5\pm
1.5^{+2.8}_{-1.7}~\mbox{MeV}$ and a width $6.1\pm
1.6^{+2.0}_{-1.4}~\mbox{MeV}$ consistent with the detector resolution.
The invariant-mass spectrum was investigated for the $pK_S^0$ and
$\bar{p}K_S^0$ separately. The measured total cross section for the
$\Theta^+$ in the kinematic range $Q^2>20~\mbox{GeV}^2$,
$p_T>0.5~\mbox{GeV}$, $|\eta|<1.5$ and $0.04<y_e<0.95$ is
$125\pm27^{+37}_{-28}~\mbox{pb}$.

A similar analysis was done by the H1 collaboration and no peak is
visible near 1520~MeV. The resulting upper limit on the $\Theta^+$
production cross section was found to be within 40 and 120 pb over the
mass range of 1.48 to 1.7 GeV and does not exclude the previously
measured cross section at ZEUS.

A search for a double strange pentaquark $\Xi^{--}_{3/2}$ as found by
the NA49 collaboration~\cite{na49} has been carried by the ZEUS
collaboration and given a negative result (see Fig.~\ref{fig:xi}).

A search for a charm pentaquark $\Theta_c$ candidate was carried out
by H1 using DIS data. The $\Theta_c$ was reconstructed via its decay
to $D^{*}p$, where $D^{*}\rightarrow D^0\pi\rightarrow K\pi\pi$. A
clear and narrow resonance (see Fig~\ref{fig:penta}(right)) is observed
for both $D^{*-}p$ and $D^{*+}\bar{p}$ with a mass of $3099\pm 3\pm
5~\mbox{MeV}$ with a width compatible with the experimental
resolution. An acceptance corrected ratio, the fraction of $D^*$
coming from the decay of the $\Theta_c$ candidate,
$R_{\mbox{corr}}(D^*p(3100)/D^*)=1.59\pm 0.33^{+0.33}_{-0.45}\%$. A
similar analysis was done by ZEUS using higher statistics and
reconstructing $D^*$ mesons both in the $K\pi\pi$ and $K\pi\pi\pi\pi$
channels. No signal near 3100 MeV is observed. ZEUS estimated the
upper limit on the acceptance corrected ratio and it is equal 0.59\%
(0.51\% for both $D^*$ decay channels) in DIS that would contradict
the measurement of H1.


\section*{References}
\begin{thebibliography}{99}

\bibitem{charm1} A. Aktas {\em et al.} [H1 Collaboration],
  \Journal{\EJP}{38}{2005}{447} [hep-ex/0408149]
 
\bibitem{charm2} S. Chekanov {\em et al.} [ZEUS Collaboration],
  \Journal{\EJP}{44}{2005}{351} [hep-ex/0508019]
 
\bibitem{charm3} S. Chekanov {\em et al.} [ZEUS Collaboration],
  contributed paper to XXIInd International Symposium on Lepton-Photon
  Interactions at High Energy, June 2005, Uppsala, Sweden

\bibitem{light1} A. Rostovtsev, In Proceedings of the 31st
  International Symposium on Multiparticle Dynamics, Datong, China,
  1-7 Sep 2001
  
\bibitem{penta1}D. Diakonov, V.Petrov, M. V. Polyakov,
  \Journal{\ZPA}{359}{305}{1997}
  
\bibitem{penta2} R. L. Jaffe and F. Wilczek,
  \Journal{\PRL}{91}{232003}{2003}
  
\bibitem{penta3}M. Karliner and H. J. Lipkin, [hep-ph/0307343];\\
  K. Cheung, [hep-ph/0308176]

\bibitem{na49} C. Alt {\em et al.} [NA49 Collaboration],
  \Journal{\PRL}{92}{042003}{2004}

\end{thebibliography}
\end{document}